\documentclass[journal=jpclcd,manuscript=article,layout=twocolumn]{achemso}

\usepackage{chemformula} % Formula subscripts using \ch{}
\usepackage[T1]{fontenc} % Use modern font encodings
\usepackage{pgfplots}
\usepackage{bm}
\usepackage[symbol]{footmisc}
\usepackage{amssymb} % provides \gtrsim

\usepackage{caption}                            %<---
\DeclareCaptionLabelFormat{Sformat}{#1 S#2}     %<---

\usepackage{mycommands}
\newcommand{\He}{\ensuremath{\hat{V}}}

\author{Daniele Furlanetto}
\affiliation[ETHZ]
{Department of Chemistry and Applied Biosciences, ETH Zürich, 8093 Zürich, Switzerland}
\email{daniele.furlanetto@phys.chem.ethz.ch}
\author{Jeremy O. Richardson}%%%%%%%%%%%%%%%%%%%%%%%%%%%%%%%
\email{jeremy.richardson@phys.chem.ethz.ch}
\affiliation[ETHZ]
{Department of Chemistry and Applied Biosciences, ETH Zürich, 8093 Zürich, Switzerland}

\title[Electronic coherences]
  {Simulating electronic coherences induced by conical intersections using MASH: Application to attosecond X-ray spectroscopy}


\begin{document}

%%%%%%%%%%%%%%%%%%%%%%%%%%%%%%%%%%%%%%%%%%%%%%%%%%%%%%%%%%%%%%%%%%%%%
%% The "tocentry" environment can be used to create an entry for the
%% graphical table of contents. It is given here as some journals
%% require that it is printed as part of the abstract page. It will
%% be automatically moved as appropriate.
%%%%%%%%%%%%%%%%%%%%%%%%%%%%%%%%%%%%%%%%%%%%%%%%%%%%%%%%%%%%%%%%%%%%%

\begin{tocentry}
    \includegraphics{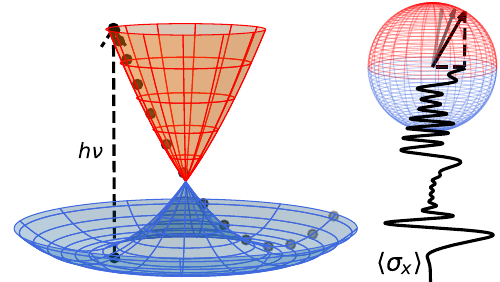}
\end{tocentry}

%%%%%%%%%%%%%%%%%%%%%%%%%%%%%%%%%%%%%%%%%%%%%%%%%%%%%%%%%%%%%%%%%%%%%
%% The abstract environment will automatically gobble the contents
%% if an abstract is not used by the target journal.
%%%%%%%%%%%%%%%%%%%%%%%%%%%%%%%%%%%%%%%%%%%%%%%%%%%%%%%%%%%%%%%%%%%%%
\begin{abstract}
  In this work, we employ trajectory-based simulations to predict the electronic coherences created by nonadiabatic dynamics near conical intersections. The mapping approach to surface hopping (MASH) is compared with standard fewest-switches surface hopping on three model systems, for which the full quantum-mechanical results are available. Electronic populations and coherences in the adiabatic representation as well as nuclear densities are computed to assess the robustness of the different methods. %The geometric-phase effects on the electronic observables are shown to be correctly simulated by the semiclassical methods.
  The results show that standard surface hopping can fail to describe the electronic coherences, whereas they are accurately captured by MASH for the same computational cost.
  In this way, MASH appears to be an excellent simulation approach for novel X-ray spectroscopies such as the recently proposed Transient Redistribution of Ultrafast Electronic Coherences in Attosecond Raman Signals (TRUECARS).
\end{abstract}

%%%%%%%%%%%%%%%%%%%%%%%%%%%%%%%%%%%%%%%%%%%%%%%%%%%%%%%%%%%%%%%%%%%%%
%% Start the main part of the manuscript here.
%%%%%%%%%%%%%%%%%%%%%%%%%%%%%%%%%%%%%%%%%%%%%%%%%%%%%%%%%%%%%%%%%%%%%

\paragraph{Introduction}
The nonadiabatic dynamics of photoexcited molecules is of fundamental interest because of their impact in chemical synthesis\cite{albini2010handbook,yoon2010visible}, biological processes\cite{vision_CIs}, and medical advancement\cite{photomedicine}. 
Ultrafast internal energy conversion and other photochemical processes are typically mediated by conical intersections (CIs), where the ground- and excited-state potential energy surfaces are degenerate.\cite{ConicalIntersections1}
Despite the fact that CIs are a theoretical construct, it has been proposed that there are physical signatures associated with them which could be measured experimentally.\cite{Muk_review_2023}
For instance, a novel X-ray spectroscopy called Transient Redistribution of Ultrafast Electronic Coherences in Attosecond Raman Signals (TRUECARS) has been designed to probe the electronic coherences generated by the transition through a CI\@.\cite{TRUECARS_intro}

Experimental spectroscopic signals are often interpreted using the help of theoretical simulations.
Whereas it has become common to calculate the electronic population dynamics using semiclassical methods such as surface hopping,\cite{Barbatti2014newtonX,Wang2016perspective,Mai2018SHARC}
it is often assumed that coherences are much harder to capture than populations.

One reason of this could be that the well-known overcoherence problem of FSSH is often tackled using decoherence corrections.\cite{Granucci2007FSSH,Granucci2010decoherence,Subotnik2011AFSSH,Jain2016AFSSH}
However, these corrections artificially modify the wavefunction coefficients, which could destroy the coherences that we wish to study.
%\footnote{We note that attempts have been made to simulate TRUECARS signals using FSSH with decoherence corrections,\cite{Mukamel_TRUECARS_FSSH} although these have not been carefully benchmarked to test whether the simulation is reliable.}
It is thus clearly preferable to avoid decoherence corrections in this case wherever possible.

Therefore (with the exception of one surface-hopping study)\cite{Mukamel_TRUECARS_FSSH} nearly all previous simulations of TRUECARS have relied on
quantum-mechanical methods either in reduced dimensionality,\cite{pnas_Mukamel}
for simplified system--bath models,\cite{TRUECARS_HEOM}
or using trajectory-guided basis functions.\cite{Mukamel2023}

The mapping approach to surface hopping (MASH),\cite{MASH,MASHreview} has recently emerged as an alternative to fewest--switches surface hopping (FSSH).\cite{Tully1990hopping,Subotnik2016review} 
In particular, it is deterministic rather than stochastic and thus avoids the problem of inconsistency between the active state and the electronic coefficients, which can plague FSSH\@.\cite{Granucci2007FSSH}
MASH is rigorously derived from a short-time limit of quantum mechanics,\cite{MASH,MASHreview} and has been shown to fix known problems of FSSH with computing nonadiabatic rates\cite{MASHrates}
or population observables starting from coherent states.\cite{Mannouch2024coherence}
Moreover, it maintains all the pros of an independent-trajectory method, allowing a first-principles on-the-fly simulation of complex molecular systems in full dimensionality.\cite{Mannouch2024MASH,cyclobutanone}
In fact, MASH can be implemented with an algorithm based on wavefunction overlaps similar to that of FSSH, although with the added benefit of being reversible and second-order in time. \cite{MASHEOM}

%In previous work, it has been shown that MASH is superior to FSSH in simulating nonadiabatic rates,\cite{MASHrates}
%and in obtaining population observables starting from coherent states.\cite{Mannouch2024coherence}
%in the present article we are doing the opposite, measuring coherences starting from population states.

In order to assess the accuracy of the MASH and FSSH methods for the evaluation of coherences, we will test them on various model systems, which describe both avoided crossings and CIs, and compare the results with fully quantum-mechanical (QM) calculations. In this way, we find MASH to be a powerful method for simulating the coherences generated in nonadiabatic processes and propose that it can be used to reliably predict TRUECARS signals in molecular systems.

%TRUECARS two (entangled) photon absorption \cite{TRUECARS_2photon}.

%TRUECARS stochastic\cite{stochastic_TRUECARS}.

\paragraph{Populations and coherences}

In this work, we focus on electronic coherences in two-state systems and describe how to calculate them with both quantum-mechanical and semiclassical methods.
Our formalism will be based on Pauli matrices defined in the adiabatic representation:\cite{MASHreview}
\begin{subequations}
\begin{align}
    \hat\sigma_x&=\ket{\phi_0}\bra{\phi_1}+\ket{\phi_1}\bra{\phi_0} \\
    \hat\sigma_y&=\iu \big(\ket{\phi_0}\bra{\phi_1}-\ket{\phi_1}\bra{\phi_0}\big) \\
    \hat\sigma_z&=\ket{\phi_1}\bra{\phi_1}-\ket{\phi_0}\bra{\phi_0}
\end{align}
\end{subequations}
where $\vert\phi_0\rangle$ and $\vert\phi_1\rangle$ are the ground and excited electronic states.
The Pauli matrices are thus implicitly dependent on the position of the nuclei, $q$.
We choose to work in the adiabatic representation because this is the natural choice for FSSH and MASH\@.
However, note that the TRUECARS signal, which we will evaluate later, is, like all physical observables, formally independent of the representation used.

In quantum mechanics, the total (nuclear and electronic) wavefunction can be expanded in the adiabatic basis as $\ket{\Psi(t)}=\ket{\chi_0(t)}\otimes\ket{\phi_0} + \ket{\chi_1(t)}\otimes\ket{\phi_1}$, although in practice we propagate it using the split-operator method in the diabatic basis and only convert to the adiabatic representation before evaluating observables. The quantum-mechanical definition of the coherences are
$\braket{\sigma_x(t)} = 2\Re\langle \chi_1(t) \vert \chi_0(t)\rangle$
and $\braket{\sigma_y(t)}=2\Im\langle \chi_1(t) \vert \chi_0(t)\rangle$,
whereas the population difference is given by $\braket{\sigma_z(t)}=\braket{\chi_1(t)|\chi_1(t)}-\braket{\chi_0(t)|\chi_0(t)}$.
It is clear that coherences require measuring the overlap of two different wavepackets, whereas the population difference is defined purely in terms of probabilities of the adiabatic states.
For this reason, it is often assumed that the populations are easier to model with semiclassical methods.
However, coherences carry phase information, which can be very important for novel spectroscopies such as TRUECARS.

In FSSH, at each trajectory corresponds an electronic wavefunction, defined as $\vert\psi(t)\rangle = c_0(t)\vert\phi_0\rangle + c_1(t)\vert\phi_1\rangle$, while the active surface, $n\in\{0,1\}$, is an additional time-dependent stochastic variable which determines the force used to evolve the nuclei.
There have been multiple suggestions for how to measure electronic observables in FSSH; we follow the procedure of using the wavefunction coefficients for the coherences and the active state $n$ to measure the populations\cite{Subotnik2016review}: $\braket{\sigma_x(t)} = \braket{2\Re c_0(t) c_1^*(t)}_\text{FSSH}$, $\braket{\sigma_y(t)} = \braket{2\Im c_0(t) c_1^*(t)}_\text{FSSH}$ and $\braket{\sigma_z(t)} = \braket{\delta_{1,n(t)} - \delta_{0,n(t)}}_\text{FSSH}$. Here, the FSSH average is taken over an ensemble of trajectories initialized with nuclear phase-space variables sampled randomly from a Wigner distribution and where the electronic wavefunction of each trajectory is initialized in the excited state, $\ket{\psi(0)}=\ket{\phi_1}$.

In MASH, the role of the electronic wavefunction is taken by a spin vector that evolves on a Bloch sphere. The nuclear force is determined by the $S_z$ component of the spin: when it is in the lower/upper hemisphere, the ground/excited-state force is used. The expectation values are written as correlation functions:
\begin{subequations} \label{eq:MASH_coh}
\begin{align}
    \braket{\sigma_x(t)} &= \big\langle 2\,h(S_z(0))\,S_x(t) \big\rangle_\text{MASH} \\
    \braket{\sigma_y(t)} &= \big\langle 2\,h(S_z(0))\,S_y(t) \big\rangle_\text{MASH} \\
    \braket{\sigma_z(t)} &= \big\langle 2|S_z(0)|\,h(S_z(0))\,\sgn(S_z(t)) \big\rangle_\text{MASH}
\end{align}
\end{subequations}
where $h$ is the Heaviside step function and the average is taken over random samples of nuclear phase-space variables from a Wigner distribution with spin vectors uniformly distributed over the Bloch sphere.\footnote[2]{Due to the factor of $h(S_z(0))$, only trajectories initialized in the northern hemisphere will contribute, and thus the sampling can be limited to that region.} As we are initializing in a population, weighting functions of $2$ or $2|S_z(0)|$ are introduced according to the MASH prescription.\cite{MASH,MASHreview}.

In both our MASH and FSSH simulations, momenta are rescaled along the direction of the nonadiabatic coupling vectors after successful hops
and reflected for all frustrated hops. This procedure is uniquely derived from the MASH formalism\cite{MASH}, while for FSSH numerous alternatives have been suggested.\cite{Mueller1997FSSH,Jasper2003FSSH,Toldo2024FSSH}

\paragraph{Avoided-crossing model}
To highlight the differences between FSSH and MASH predictions of coherences, we utilize a system of coupled one-dimensional harmonic oscillators exhibiting an avoided crossing, which is described by the following diabatic potential (whose eigenvectors are the adiabatic states):
\begin{align}
    \He = \thalf m\omega^2 q^2 + \begin{pmatrix}
        \kappa q + \varepsilon & \Delta \\
        \Delta & - \kappa q - \varepsilon\\
    \end{pmatrix}
    \label{eq:avoided_crossing_potential}
\end{align}
with the parameters defined in the SI.

\begin{figure*}
    \centering
    \includegraphics{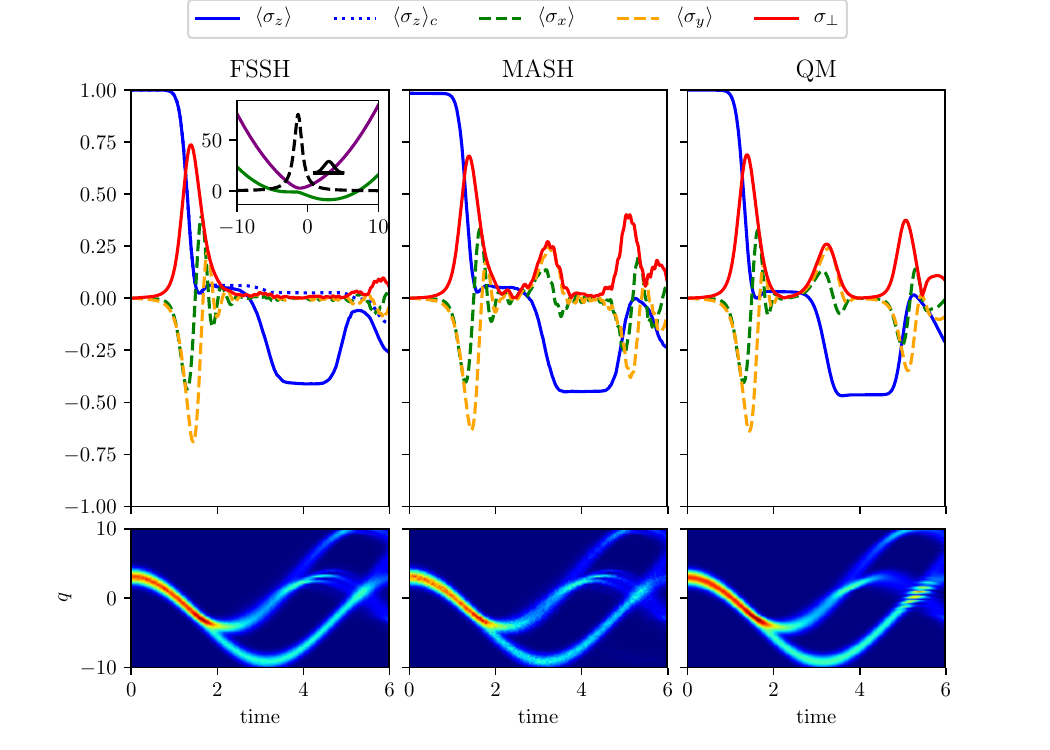}
    \caption{The first row presents the adiabatic populations and coherences calculated by semiclassical and quantum methods for the avoided-crossing model. The alternative FSSH measure of $\braket{\sigma_z(t)}_c = \braket{|c_1(t)|^2-|c_0(t)|^2}_\mathrm{FSSH}$ is also shown to illustrate the inconsistency with $\braket{\sigma_z(t)}$. $\sigma_\perp=\sqrt{\braket{\sigma_x}^2 + \braket{\sigma_y}^2}$ is the absolute value of the coherences. The inset shows the adiabatic energies, the nonadiabatic coupling, and the initial wavepacket. The second row shows the time-dependent nuclear density.}
    \label{fig:harm_pop_coh}
\end{figure*}

Figure~\ref{fig:harm_pop_coh} shows that both semiclassical methods correctly describe the population evolution, despite the fact that FSSH exhibits a large inconsistency error, as seen by the discrepancy between the average populations measured by the wavefunction coefficients and the fraction of trajectories on the given state. The nuclear-density evolution also shows comparable results for all methods [Fig.~\ref{fig:harm_pop_coh}]. The only significant difference is that the full QM simulation shows signs of nuclear interferences near $t = 5$. % between wavepackets that jumped to the ground state at different previous times.
This is the only feature of the simulations that cannot be captured by MASH. However, it does not affect the quality of the electronic coherences, which are in excellent agreement at all times.

In contrast, FSSH fails completely to describe the coherences induced by the second and third passages through the coupling region ($t\approx3$ and $t\approx5$). This is a direct consequence of the inconsistency problem in FSSH, which is fixed in MASH.
% Neither of those coherences are decribed by FSSH, while they are correctly detected by MASH. 
This is a clear numerical example of the benefits of the rigorous MASH prescription over the traditional FSSH approach.
Similar behaviour was observed for Tully's third model in Ref.~\citenum{MASH}.

\paragraph{Jahn--Teller models}
% - Intro
In two or more dimensions, it is possible (and quite common) to encounter conical intersections. CIs are known to play a key role in photochemical reactions\cite{ConicalIntersections1} and it is therefore of great interest to study the dynamics in their vicinity.
For this work, we study the coherences generated on passing through a conical intersection.
One might assume that due to the huge nonadiabatic coupling in this region, the coherence signals will be large. However, if the nuclear wavepacket is spread on both sides of the CI, the paths that pass by the CI on different sides will experience opposite nonadiabatic couplings, leading to total destructive interference of the coherences\cite{symmetry_TRUECARS, symmetry_coherences, Kowalewski2022}.

% - Model
To illustrate this behaviour we used a simple symmetric Jahn--Teller model described by the following diabatic potential:
\begin{align}
    \He = \thalf m\omega^2(q_1^2 + q_2^2) + \begin{pmatrix}
        \kappa q_1 & \lambda q_2 \\
        \lambda q_2 & - \kappa q_1 \\
    \end{pmatrix}
    \label{eq:CI_potential}
\end{align}
%
% - Antisymmetric evolution causes zero coherences in QM
Solving for the adiabatic energies (as the eigenvalues of this matrix) leads to a peaked CI at the origin.
The initial wavepacket is chosen to be a Gaussian on the excited state, centered at positive $q_1$ with zero average momentum; the initial conditions are thus symmetric with respect to reflection in the $q_2$ axis.
As the symmetric wavepacket passes near the CI, it generates an antisymmetric wavepacket on the ground state
due to the antisymmetric coupling $d\cdot p$, where $d=\braket{\phi_1|\pder{}{q}\phi_0}$ is the nonadiabatic coupling and $p$ is the momentum.
The coherences $\braket{\sigma_x}$ and $\braket{\sigma_y}$ will thus be zero since the overlap $\braket{\chi_1(t)|\chi_0(t)}$ vanishes due to the different symmetries of the wavefunctions.\footnote[3]{Note that the diabatic coherences are also zero by similar symmetry arguments.}
%\cite{symmetry_coherences, symmetry_TRUECARS}

% - Geometric phase causes zero coherence in MASH
The MASH coherences are also zero by symmetry. %In the case of symmetric initial conditions, 
For every trajectory starting with initial conditions $(q_1,q_2,p_1,p_2,S_x,S_y,S_z)$ there is another trajectory with the same weight starting with $(q_1,-q_2,p_1,-p_2,-S_x,-S_y,S_z)$. These two trajectories will experience equivalent forces, but opposite coupling $d\cdot p$, leading to antisymmetric evolution of the spin vector, and therefore zero electronic coherences on average.
For a similar reason, fully converged FSSH simulations also give zero coherences.
%This behaviour can be ensured even for a finite number of trajectories through a symmetrized initial sampling (see SI). Using the same argument, the FSSH coherences are also zero in the limit of infinite trajectories, but they cannot be ensured to be zero for a finite number because of the stochastic FSSH algorithm.

\begin{figure}
    \centering
    \includegraphics{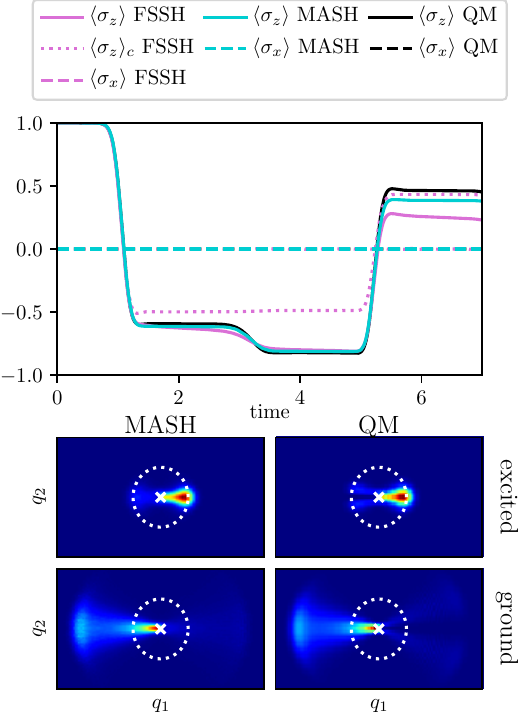}
    \caption{Upper: Time-dependent populations and coherences for the Jahn--Teller model with symmetric initial conditions. Note that coherences are zero in all methods. Lower: MASH and QM nuclear densities on the excited and ground states, averaged over the simulation time.
    The white dotted circle indicates the degenerate ground-state minimum and the white cross indicates the CI.}
    \label{fig:CI_plot_symm}
\end{figure}

% - Comment population of symmetric populations
The numerical results in Figure~\ref{fig:CI_plot_symm} confirm that, as expected, the coherences are zero throughout the entire dynamics.
Nonetheless, the populations show that nonadiabatic transitions are taking place.
The FSSH population is clearly affected by inconsistency with the electronic coefficients and leads to poor predictions after the third crossing region, whereas MASH is significantly more accurate.

%To describe the MASH and FSSH coherences it is useful to discuss the geometric phase effect. The adiabatic evolution of the electronic wavefunction around a CI leads to the formation of a phase, that results in a sign change after a complete rotation around the CI\@.\cite{Mead1992geometric} It is clear that MASH and FSSH correctly describe the geometric phase in the sense that, by construction, their dynamics reproduces the time-dependent Schr\"odinger equation for the electronic states along a given nuclear trajectory.
%The geometric phase leads to a suppression of the coherences for symmetric initial conditions: for every trajectory starting with initial conditions $(q_1,q_2,p_1,p_2,S_x,S_y,S_z)$ there is exactly one other trajectory starting with $(q_1,-q_2,p_1,-p_2,-S_x,-S_y,S_z)$ that will have symmetric nuclear evolution, but opposite electronic evolution, leading to opposite signs of $S_x$ and $S_y$, that will average to zero. In this case, the coherences are suppressed not because of the presence of an antisymmetric nuclear wavefunctions, but because of the geometric phase in the electronic wavefunction.\cite{geometric_phase_CI}

% - Geometric phase causes node in density for QM, not for MASH
The geometric-phase effect\cite{Mead1992geometric} influences the nuclear wavefunction by enforcing a node around the CI. This effect can be observed in the quantum nuclear density plotted in Fig.~\ref{fig:CI_plot_symm}\cite{exat_factorization, Valahu:2022kfg}.
In MASH and FSSH, the nuclear densities do not exhibit a node because they do not capture nuclear interferences.
It is thus clear that the suppression of coherences is caused by the symmetry of the problem\cite{symmetry_coherences} and is not directly a geometric-phase effect as has been suggested in the literature\cite{symmetry_TRUECARS}.

%\blue{%Maybe write a separate paragraph about the geometric phase here...
%Discussions of dynamics at CIs are inextricably linked to the geometric-phase effect. 
%On one hand, it is clear that MASH and FSSH correctly describe the geometric phase in the sense that, by construction, their dynamics reproduces the time-dependent Schr\"odinger equation for the electronic states along a given nuclear trajectory.
%They thus naturally capture the fact that the adiabatic states change sign when winding around the CI\@.\cite{Mead1992geometric}
%On the other hand, %this global phase has no effect on the physical observables
%as they do not capture nuclear interference, there is no discernible effect on the physical observables.
%Therefore, whereas the quantum-mechanical nuclear density exhibits a node due to cancellation between wavepackets of opposite sign,\cite{Fede+Izmaylov+etc, Valahu:2022kfg} the semiclassical methods do not [Fig.~\ref{fig:CI_plot_symm_den}].
%Note that semiclassical methods such as MASH and FSSH cannot be expected to capture nuclear quantum effects such as the vibrational excitation that arises from the Berry phase on the electronic wavefunction.
%For this reason, the semiclassical dynamics do not present nodes in the nuclear density, while the full QM dynamics shows the presence of nodes caused by the Berry phase [Fig.\ref{fig:CI_plot_symm_den}].
%}

\begin{figure}
    \centering
    \includegraphics{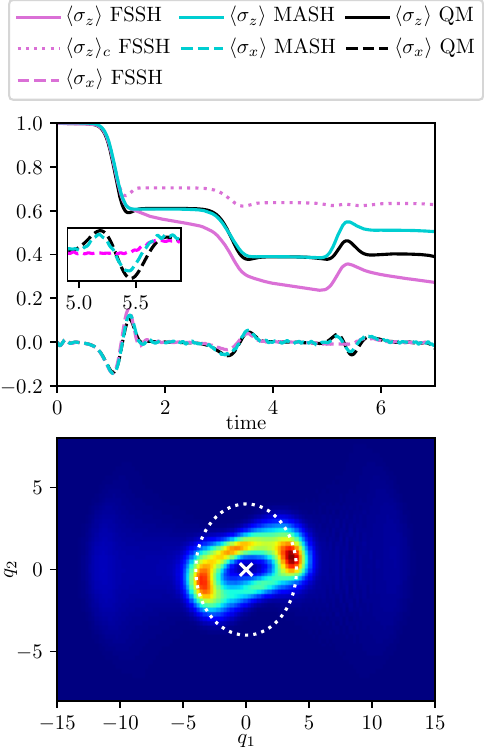}
    \caption{Upper: populations and coherences for the Jahn--Teller model with asymmetric initial conditions. The inset shows the coherences around the third crossing $t \approx 5.5$. Lower: QM nuclear density averaged over the simulation time. MASH and FSSH give almost identical results (see SI).}
    \label{fig:CI_plot_p}
\end{figure}

% - Comment the antisymmetric case
Starting the wavepacket with a non-zero average initial momentum in the direction of $q_2$ breaks the symmetry. The wavepacket orbits around the CI, resulting in a dynamics that can be interpreted as multiple passages through avoided crossing regions. The nuclear density [Fig.~\ref{fig:CI_plot_p}] evolves similarly for all methods. The electron population dynamics is best captured by MASH, whereas FSSH is again strongly affected by internal inconsistency. The electron coherence dynamics is captured by all methods in the first and second crossings, but FSSH fails completely in the third crossing ($t\approx 5.5$). There is no obvious effect of the geometric phase here since the whole wavepacket goes around the CI on one side such that there is no nuclear interference.

\paragraph{TRUECARS signal}
The TRUECARS experiment was proposed to study the passage of a wavepacket through a region of nonadiabatic coupling, such as near a CI\cite{TRUECARS_intro}. The experimental observable is the intensity of the scattered Raman signal of a hybrid X-ray probe pulse  after an optical pump has initialized the system in an excited-state population. The relevant response function\cite{mukamel1995principles} is the time-dependent expectation value of the polarizability $\langle\alpha(t)\rangle$. The spectroscopic signal can then be simulated as a function of the time delay $T$ and of the Raman shift $\omega$ as
\begin{align}\label{eq:Sfd}
    \mathcal{S}(\omega,T)=2\Im\hspace{-0.3em}\int_{-\infty}^{+\infty}\hspace{-1.6em} \mathrm{d}t\, \eu{\iu\omega(t-T)}\mathcal{E}_\mathrm{B}^*(\omega)\mathcal{E}_\mathrm{N}(t\!-\!T) \langle\alpha(t)\rangle
\end{align}
where $\mathcal{E}_\mathrm{B}(\omega)$ and $\mathcal{E}_\mathrm{N}(t)$ are, respectively, the broadband and narrowband probe-pulse envelopes. %These envelopes are Gaussians in the respective frequency and time-domain.
Finally, an alternative way of visualizing the results is offered by the Wigner spectrogram:\cite{pnas_Mukamel} %Given a fixed value for the frequency $\omega_\mathrm{R}$, it is possible to obtain the Wigner spectrogram: 
\begin{align}\label{eq:Wigner_spectr}
    \mathcal{W}(T,\omega)=\hspace{-0.3em}\int_{-\infty}^{+\infty}\hspace{-1.6em} \mathrm{d}t\, \mathcal{S}(\omega_\mathrm{R}, T\!+\!\tfrac{t}{2})\mathcal{S}(\omega_\mathrm{R}, T\!-\!\tfrac{t}{2}) \eu{-\iu\omega t}
\end{align}
which is defined using a fixed value for the frequency $\omega_\mathrm{R}$.

The computation of $\mathcal{S}$ and $\mathcal{W}$ are just simple post-processing operations.
All that is required is the expectation value $\braket{\alpha(t)}$, which can be computed by QM, MASH or FSSH simulations.
For instance, in quantum mechanics, $\langle\alpha(t)\rangle = \langle\Psi(t)\vert\hat{\alpha}\vert\Psi(t)\rangle$.
FSSH and MASH are formulated in the adiabatic basis, and thus it is useful to decompose the real-valued polarizability operator into the basis of adiabatic Pauli matrices as:
\begin{align}
    \hat{\alpha}(q) = \alpha_z(q)\hat{\sigma}_z + \alpha_x(q)\hat{\sigma}_x
    \label{eq:alpha_adiab}
\end{align}
In general the polarizability operator is a tensor in the spatial coordinates.  Here, we focus on the scalar isotropic component, although the generalization is trivial. Using Eq.~\eqref{eq:alpha_adiab}, the FSSH definition of adiabatic populations and coherences, and the MASH coherences [Eq.~\eqref{eq:MASH_coh}] we obtain the expectation values:
\begin{subequations}
\begin{align}
    \begin{split}
        \langle\alpha(t)\rangle =& \big\langle\alpha_z(q(t))(\delta_{1,n(t)}-\delta_{0,n(t)}) \\
        &+ 2\alpha_x(q(t)) \Re[c_{1}^*(t)c_{0}(t)] \big\rangle_\text{FSSH}
    \end{split}\\
    \begin{split}
        \langle\alpha(t)\rangle =& \big\langle 2|S_z(0)|h(S_z(0))\alpha_z(q(t))\sgn(S_z(t)) \\
        &+ 2h(S_z(0))\alpha_x(q(t))S_x(t) \big\rangle_\text{MASH}
    \end{split}
\end{align}
\end{subequations}
The polarizability thus has contributions from both the populations and the coherences, although the coherences are expected to dominate the TRUECARS signal as the Fourier transform of Eq.~\eqref{eq:Sfd} will pick out oscillating terms.

\paragraph{Model of acrolein}
To test the reliability of semiclassical methods for simulating TRUECARS observables, we used the nonlinear two-dimensional parametrized model of acrolein from Ref.~\citenum{TRUECARS_intro}, with minor modifications to ensure that the potential is bounded (see SI). As in the original work, the dynamics are initialized in the excited adiabatic state, with a Gaussian wavepacket with zero average momentum.

\begin{figure}
    \centering
    \includegraphics{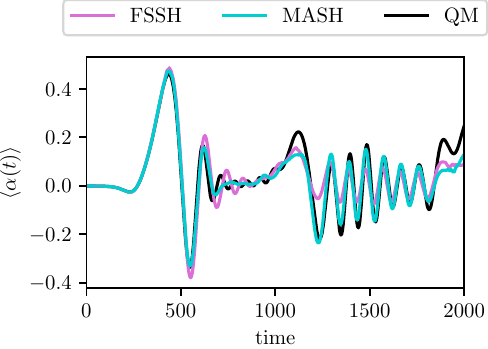}
    \caption{Time-evolution of the polarizability expectation value for the model system of acrolein.}
    \label{fig:MukII_CI_a}
\end{figure}

\begin{figure*}
    \centering
    \includegraphics{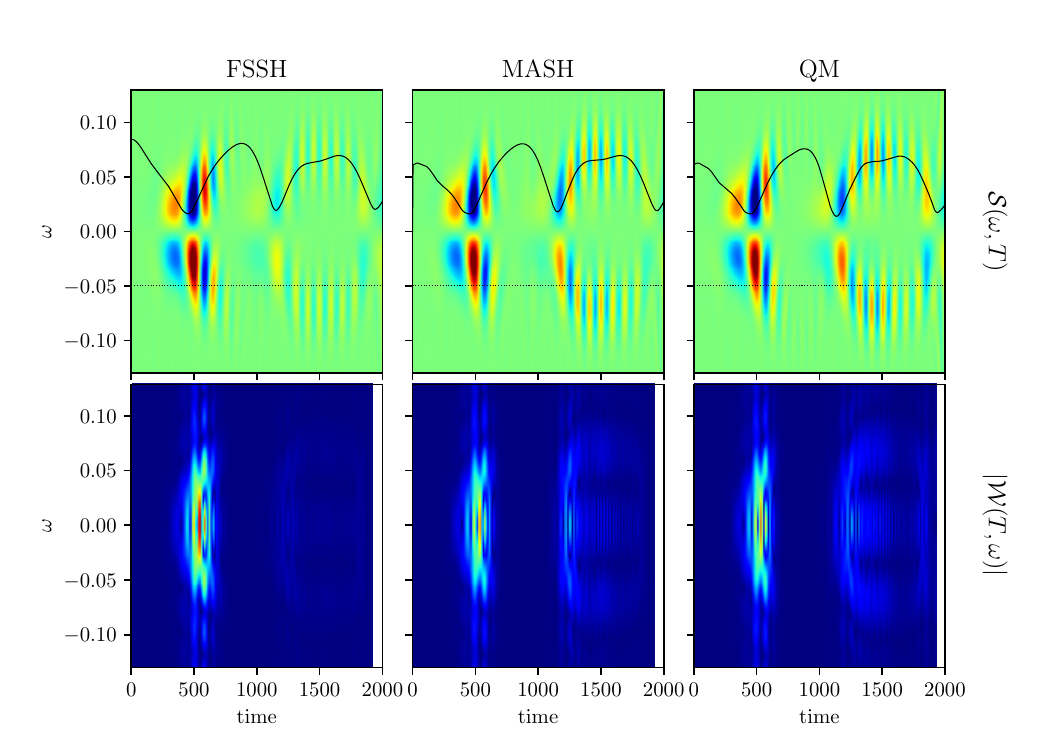}
    \caption{Upper: spectroscopic signal $\mathcal{S}(\omega, T)$ overlaid with $\bar{\omega}(T)$. Lower: absolute value of Wigner spectrogram $|\mathcal{W}(T, \omega)|$ for the reference frequency $\omega_\mathrm{R}$ shown by a dotted line in the upper panels. All quantities are in atomic units.}
    \label{fig:MukII_CI_SW}
\end{figure*}

\begin{figure}
    \centering
    \includegraphics{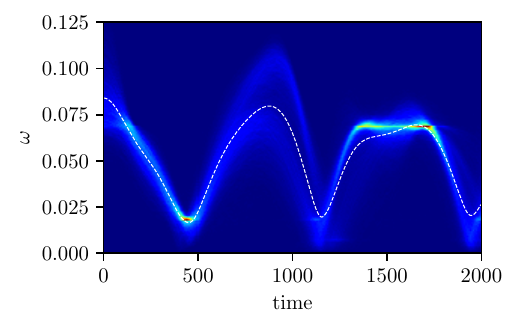}
    \caption{Energy shift between the adiabatic states $\Delta V$, measured from the ensemble of MASH trajectories weighted by the corresponding coherence factor $\sqrt{S_x(t)^2 + S_y(t)^2}$. The time-dependent mean frequency $\bar{\omega}(t)$ is overlaid as a dashed line.}
    \label{fig:DV_main}
\end{figure}

Figure~\ref{fig:MukII_CI_a} shows the time evolution of the polarizability, for which the MASH result closely resembles the QM result.
%MASH therefore accurately predicts the TRUECARS signal [Fig.~\ref{fig:MukII_CI_SW}].
However, FSSH makes significant errors. In particular, the amplitude of the oscillations are underestimated after about $t \approx 1200$.
This is a consequence of its poor description of the coherences (see SI).

Postprocessing the time-dependent polarizability from each method gives the TRUECARS signals, $\mathcal{S}(\omega,T)$, shown in Fig.~\ref{fig:MukII_CI_SW}.
MASH gives an accurate prediction, whereas FSSH shows much weaker coherences at longer times.
The error of FSSH is also clearly highlighted by the Wigner spectrogram, where the signal is almost completely washed out.

%\blue{
In addition to predicting the TRUECARS signal, the ensemble of MASH trajectories can be used to interpret it.
In order to do this, we adapt an idea from Ref.~\citenum{TRUECARS_intro} to relate the envelope of the signal to the time-dependent adiabatic energy gaps $\Delta V(q) = V_1(q)-V_0(q)$.
%In particular, we estimate the envelope as 
%\begin{align}
%    P(\omega,t) &\approx \Braket{2\,h(S_z(0)) S_\perp(t) \delta(\omega_t-\omega) \mathcal{E}_\mathrm{B}(\omega_t)}_\text{MASH}
%\end{align}
%where $S_\perp(t)=\sqrt{S_x^{2}(t) + S_y^{2}(t)}$ and $\omega_t=\Delta V(q(t))$.
A justification for this connection is presented in the SI.

In Fig.~\ref{fig:DV_main}, we present a time-dependent histogram of $\omega(t)=\Delta V(q(t))$ from the ensemble of MASH trajectories weighted by $2h(S_z(0))\sqrt{S_x^{2}(t) + S_y^{2}(t)}$, whose
shape roughly follows that of the envelope of $\mathcal{S}(\omega,T)$.
For comparison, we also plot the mean frequency of this distribution $\bar{\omega}(t)$ on both Fig.~\ref{fig:MukII_CI_SW} and Fig.~\ref{fig:DV_main}.
It is clear from the minima in $\bar{\omega}(t)$ that the trajectories pass through a region of strong nonadiabatic coupling (where the energy gap is small) three times. %, highlighted as minima of $\bar{\omega}(T)$.
Moreover, after the second crossing a high proportion of the trajectories find themselves in regions with similar potential-energy gaps ($\Delta V\approx0.07$).  The spin vectors of this set of trajectories thus evolve in phase and is what leads to the long-lasting coherences seen in the polarizability.
%This behaviour is confirmed by a visualization of the trajectories (see SI), which is simple to make in this low-dimensional problem.
%The advantage of trajectory-based methods is that they can also be easily analysed in a similar way for high-dimensional systems.
%The mean frequency $\bar{\omega}(T)$ can be used to interpret the crossing through non-adiabatic coupling regions in systems with more degrees of freedom.
%Note that in more complex problems involving multiple CIs,\cite{Mukamel2023} it may also be useful to look at the terms generated by selected sets of trajectories categorized by their similarity.

%\blue{
The Wigner spectrogram shows a strong feature at $T\approx 500$ corresponding to the creation of coherence during the first nonadiabatic transition.  Its spread in the range $\omega\approx[-0.06,+0.06]$ represents the potential-energy splitting in this region and includes a significant contribution from the conical intersection degeneracy at $\omega=0$.
At longer times $T\gtrsim 1200$,
we observe three horizontal stripes in the Wigner spectrogram at $\omega\approx-0.07$, 0, and $+0.07$.
Ref.~\citenum{pnas_Mukamel} interpreted similar features in terms of the dominating vibronic eigenstates which contribute to the coherence and claimed that a classical treatment could not describe such time-independent quantum states.
However, our MASH simulation seems to capture the behaviour almost perfectly, despite being based on an ensemble of classical trajectories with time-dependent energy splittings.
This implies that the horizontal features may not be signatures of quantum-mechanical vibronic eigenstates at all and rather just electronic coherences from different parts of the trajectory ensemble.
%}

%The dips in $bar{\omega}(T)$ shown in Fig.~\ref{fig:MukII_CI_a} clearly indicate that the MASH trajectories pass through a region of strong nonadiabatic coupling three times.
%This is confirmed by visualizations of the swarm of trajectories in the SI\@.
%The similarity between the numerical values of $\bar{\omega}(T)$ demonstrates that MASH can give a similar level of insight into the dynamics in the vicinity of a conical intersection as a full quantum-mechanical simulation.
%In more complex problems involving multiple CIs,\cite{Mukamel2023} it may also be useful to look at the terms generated by selected sets of trajectories categorized by their similarity.

\paragraph{Conclusion}
Our results show that it is possible to accurately calculate observables based on electronic coherences using semiclassical MASH dynamics, despite the fact that nuclear quantum effects are neglected. 
In contrast, the FSSH method fails not because of the semiclassical approximation \emph{per se}, but because of its well-known overcoherence problem leading to its inconsistency error.
Even in cases where FSSH gives reasonable predictions of the populations, it can fail dramatically for the coherences.
This is related to the fact that more reliable population results are obtained from active surfaces than from electronic coefficients, but that the coherences have to be computed from electronic coefficients.
MASH, on the other hand, does not suffer from the inconsistency error and can predict accurate populations and coherences from its ensemble of classical trajectories.

Although we have focused on two-state systems here, MASH has been recently generalized to multi-state systems.
For photochemistry, we recommend the approach of Ref.~\citenum{unSMASH}, whereas for exciton models, Refs.~\citenum{Runeson2023MASH} and \citenum{Runeson2024MASH} are preferred.\cite{MASHreview}
Additionally, as the MASH algorithm has the same computational cost as FSSH, it can also be used with on-the-fly electronic-structure calculations.\cite{cyclobutanone,cyclobutanone_Johan,Mannouch2024MASH}
This will allow reliable ab initio simulations of coherences in molecules treated in full-dimensionality.
In this way we can provide mechanistic understanding and theoretical support to novel nonlinear spectroscopies such as TRUECARS.

%%%%%%%%%%%%%%%%%%%%%%%%%%%%%%%%%%%%%%%%%%%%%%%%%%%%%%%%%%%%%%%%%%%%%
%% The "Acknowledgement" section can be given in all manuscript
%% classes.  This should be given within the "acknowledgement"
%% environment, which will make the correct section or running title.
%%%%%%%%%%%%%%%%%%%%%%%%%%%%%%%%%%%%%%%%%%%%%%%%%%%%%%%%%%%%%%%%%%%%%
\begin{acknowledgement}
The authors thank Marco Garavelli for introducing them to TRUECARS.
\end{acknowledgement}

%%%%%%%%%%%%%%%%%%%%%%%%%%%%%%%%%%%%%%%%%%%%%%%%%%%%%%%%%%%%%%%%%%%%%
%% The same is true for Supporting Information, which should use the
%% suppinfo environment.
%%%%%%%%%%%%%%%%%%%%%%%%%%%%%%%%%%%%%%%%%%%%%%%%%%%%%%%%%%%%%%%%%%%%%
\begin{suppinfo}

The system parameters, the initial conditions for the simulations and the definition of the post-processing procedure are defined in the Supporting Information. We also report additional results of the time-dependent polarizability and coherences, and the symmetry treatment of the Jahn--Teller model.

\end{suppinfo}

%%%%%%%%%%%%%%%%%%%%%%%%%%%%%%%%%%%%%%%%%%%%%%%%%%%%%%%%%%%%%%%%%%%%%
%% The appropriate \bibliography command should be placed here.
%% Notice that the class file automatically sets \bibliographystyle
%% and also names the section correctly.
%%%%%%%%%%%%%%%%%%%%%%%%%%%%%%%%%%%%%%%%%%%%%%%%%%%%%%%%%%%%%%%%%%%%%
\bibliography{references,sample} % please do not edit references.bib (as it will be automatically updated), so add your own references to sample.bib

\onecolumn
%\counterwithin*{figure}{section}
%\let\svsection\section
\setcounter{figure}{0}
\renewcommand\thefigure{S\arabic{figure}}
\renewcommand\thetable{S\arabic{figure}}
\section*{Supporting Information}

\paragraph{Time-dependent polarizability}
We only need to treat the traceless part of the polarizability operator $\hat{\alpha}$, since an identity component of this operator gives zero signal in TRUECARS\@.  Note also that the polarizability is real as the Hamiltonian is real. We will additionally choose the diagonal contributions to the diabatic polarizability to be zero, as suggested in Ref.~\citenum{TRUECARS_intro}, as they only contribute to the signal during the nonadiabatic passage and not to the long-lived coherences. Moreover we choose the diabatic polarizability to be independent of nuclear position and note that the proportionality constant is unimportant.
Thus, in the \emph{diabatic} representation, the polarizability operator is
\begin{align}
    \hat{\alpha}(q) = \begin{pmatrix} 0 & 1 \\ 1 & 0 \end{pmatrix}
\end{align}
for all our models.

The \emph{adiabatic} components of the polarizability can be obtained using the adiabatic Pauli matrices (which are implicitly $q$-dependent):
\begin{subequations}
\begin{align}
    \alpha_x(q)&=\thalf\Tr[\hat{\alpha}(q)\hat{\sigma}_x]\\
    \alpha_z(q)&=\thalf\Tr[\hat{\alpha}(q)\hat{\sigma}_z]
\end{align}
\end{subequations}
Note that the $\alpha_y(q)$ component of the polarizability is zero since we assume the operator $\hat{\alpha}(q)$ to be real. %The identity component of the $\hat{\alpha}(q)$ can be neglected since it produces a constant value that does not contribute to the TRUECARS signal $\mathcal{S}$.
Note that even if the diabatic polarizability is constant, the adiabatic polarizability will depend on the nuclear position through the Pauli matrices.
The polarizability of real molecules would have tensorial spatial parts; each component can be treated in exactly the same way as above.

\paragraph{Avoided-crossing model}
The avoided-crossing model described in Eq.~\eqref{eq:avoided_crossing_potential} of the main paper is defined using the following parameters in reduced units:  $\hbar=1$, $m = 1$, $\omega = 1$, $\kappa = 3$, $\varepsilon = 4$, $\Delta = 2$.

The initial wavefunction on the excited adiabatic state is described by a Gaussian wavepacket (centred at $q=\kappa/(m\omega^2)$ and $p=0$ with a standard deviation that corresponds to $\omega$).
The semiclassical trajectories are sampled from the corresponding Wigner distribution.
\begin{subequations}
\begin{align}
    \chi_1(q,t=0) &= \sqrt[4]{\frac{m\omega}{\pi\hbar}}\,\eu{-\frac{m\omega}{2\hbar}(q-\frac{\kappa}{m\omega^2})^2} \\
    \rho_1(q, p ,t=0) &= \frac{1}{\pi\hbar}\,\eu{-\frac{m\omega}{\hbar}(q-\frac{\kappa}{m\omega^2})^2}\,\eu{-\frac{p^2}{m\hbar\omega}}
\end{align}
\end{subequations}
These initial conditions model an instantaneous excitation at the Franck--Condon point.

\begin{figure}
    \centering
    \includegraphics{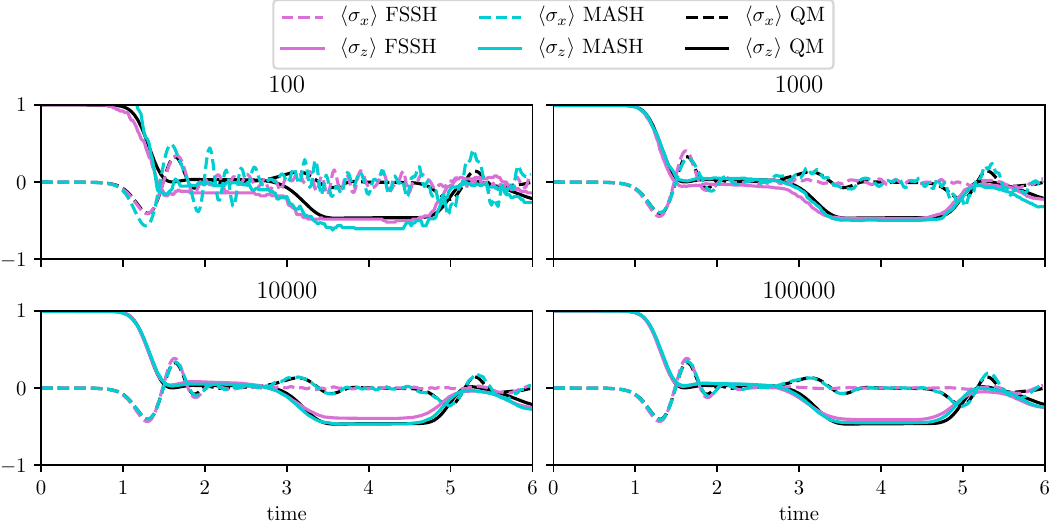}
    \caption{MASH and FSSH populations and coherences averaged over different number of trajectories for the avoided-crossing model.}
    \label{fig:convergence}
\end{figure}

Throughout the entire paper, the nuclear dynamics are integrated using the velocity-Verlet algorithm, while the electronic dynamics are integrated using the nonadiabatic coupling. For MASH the electronic integration is symmetrized and bisected 10 times in the jumping steps\cite{MASHEOM}. The simulation timestep is $\delta t = 0.02$, which ensures convergence of the trajectory integration. For MASH and FSSH the convergence of the ensemble has been ensured using $10^5$ trajectories.  However, in practice it is not necessary to use so many. In Figure \ref{fig:convergence} the results are compared for different number of trajectories, showing that MASH and FSSH converge at a similar rate and that in both cases, reasonable results can already be obtained using about 1000 trajectories.

To facilitate the convergence of the coherences at short times, the MASH spin-vector is sampled in a symmetric manner with respect to rotations about the $z$-axis. In practice for every trajectory with initial conditions $(q,p,S_x,S_y,S_z)$, another trajectory is initialized with $(q,p,-S_x,-S_y,S_z)$. This approach speeds up the convergence of the results (at least at short times). 

\begin{figure}
    \centering
    \includegraphics{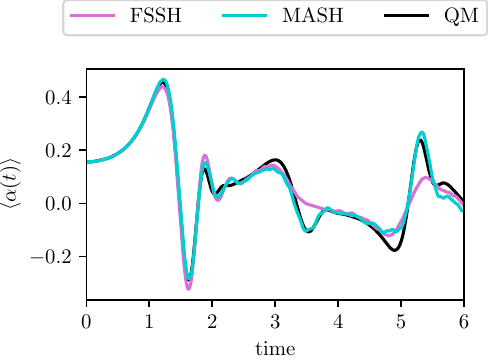}
    \caption{Time-evolution of the polarizability expectation value for the avoided-crossing model.}
    \label{fig:harm_alpha}
\end{figure}

Figure \ref{fig:harm_alpha} shows the time-dependent expectation value of the polarizability obtained from the simulations described in the main text.
As a consequence of its error in computing adiabatic coherences, the FSSH polarizability shows significant differences from the full quantum-mechanical (QM) results, especially near the peaks at $t \approx 3.5$, $5$ and $5.3$. In contrast, the MASH prediction is in much better agreement (although the peak at $t\approx 5$ still shows a small discrepancy).

\paragraph{2D Jahn--Teller models}
The symmetric Jahn--Teller model described in Eq.~\eqref{eq:CI_potential} of the main text is defined using the following parameters: $\hbar=1$, $m = 1$, $\omega = 1$, $\kappa = \lambda = 4$ in reduced units. The initial wavefunction is described by a Gaussian wavepacket in the excited adiabatic state which is centered in position $\bar{q} = (\kappa/m\omega^2, 0)$ with a standard deviation that corresponds to $\omega$ in both directions. %The initial average position is arbitrary, it does not represents a Franck-Condon excitation.
The average initial momentum $\bar{p}=(0,0)$ in the first simulation, while it is $\bar{p} = (0, 2)$ in the second case. Notice that non-symmetric initial conditions can be obtained setting $\kappa\neq\lambda$ and shifting the initial wavepacket. In both cases, the semiclassical trajectories are sampled from the corresponding Wigner distributions.
\begin{subequations}
\begin{align}
    \chi_1(q,t=0) &= \sqrt{\frac{m\omega}{\pi\hbar}}\,\eu{-\frac{m\omega}{2\hbar}(q_1-\bar{q}_1)^2}\,\eu{-\frac{m\omega}{2\hbar}q_2^2+\frac{\iu}{\hbar} \bar{p}_2 q_2} \\
    \rho_1(q_1, q_2, p_1, p_2, t=0) &= \frac{1}{\pi^2\hbar^2}\,\eu{-\frac{m\omega}{\hbar}(q_1-\bar{q}_1)^2}\,\eu{-\frac{m\omega}{\hbar} q_2^2}\,\eu{-\frac{p_1^2}{m\hbar\omega}}\,\eu{-\frac{(p_2-\bar{p}_2)^2}{m\hbar\omega}}
\end{align}
\end{subequations}

The simulation timestep is $\delta t = 0.01$. For MASH and FSSH the convergence has been ensured using $10^5$ trajectories. In the case of the first simulation, the trajectory sampling is symmetrized in the nuclear phase-space for both MASH and FSSH: for each trajectory with initial conditions $(q_1,q_2,p_1,p_2,S_x,S_y,S_z)$, another trajectory is initialized with $(q_1,-q_2,p_1,-p_2,-S_x,-S_y,S_z)$. Using this symmetrized approach, the MASH coherences are exactly zero at each time. The FSSH coherences are not ensured to be zero (except in the limit of infinite sampling), since the random numbers for the hopping algorithm are not symmetrized.

\begin{figure}
    \centering
    \includegraphics{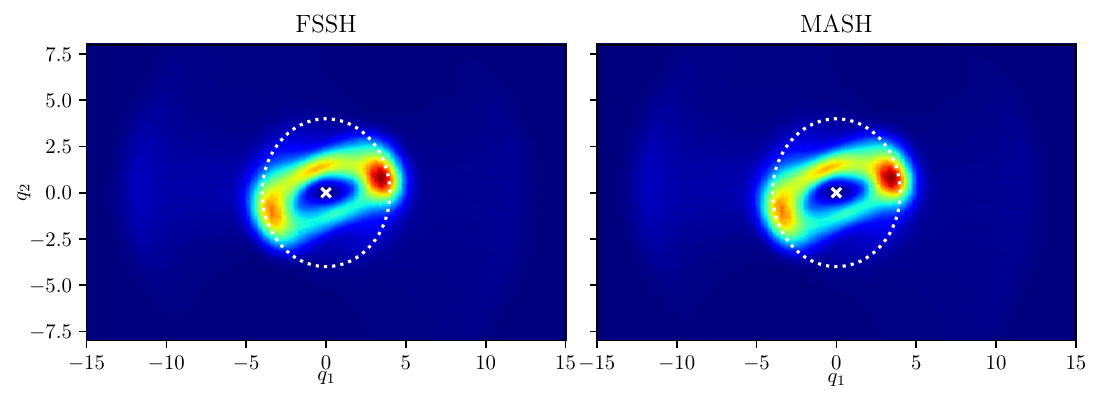}
    \caption{MASH and FSSH nuclear density averaged over the simulation time, both indistinguishable from the QM result given in Fig.~\ref{fig:CI_plot_p} of the main text.}
    \label{fig:CI_si_p}
\end{figure}

\begin{figure*}
    \centering
    \includegraphics{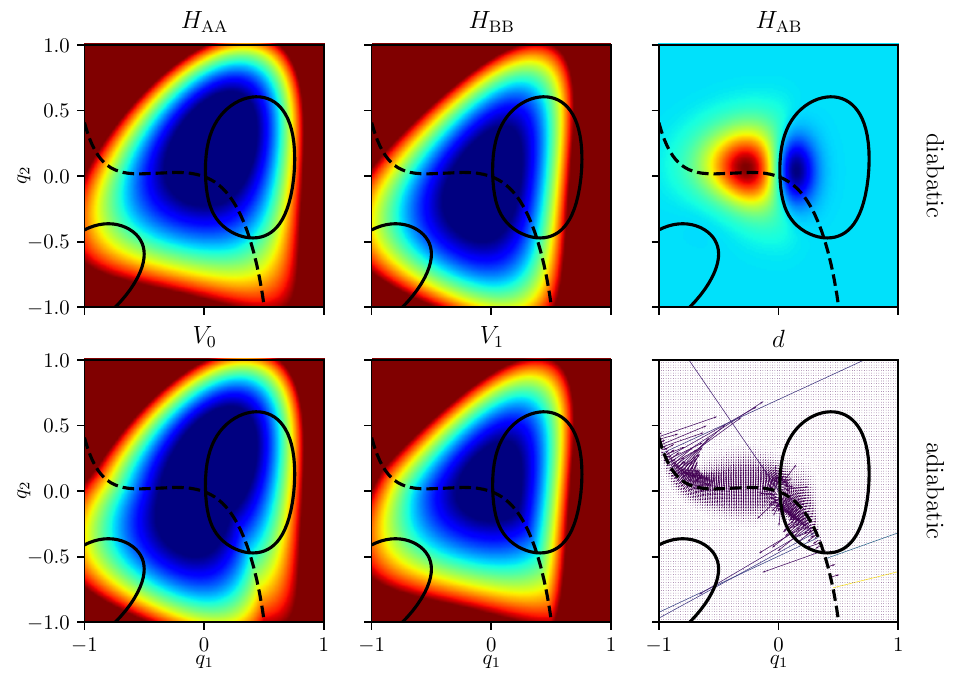}
    \caption{Diabatic and adiabatic energies, diabatic coupling and nonadiabatic coupling vectors, $d$, for the model of acrolein.}
    \label{fig:2DMuk_pot}
\end{figure*}

\begin{figure}
    \centering
    \includegraphics{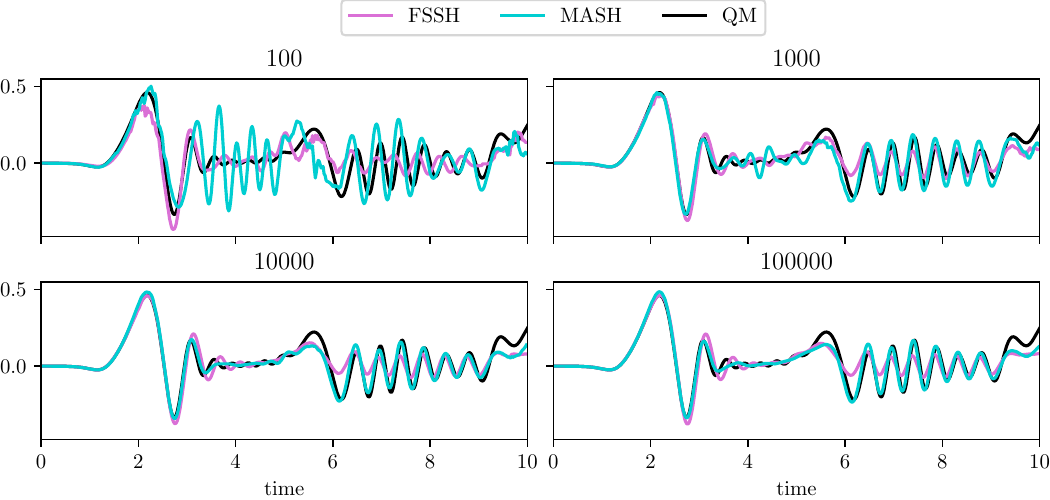}
    \caption{MASH and FSSH polarizabilities averaged over different number of trajectories for the model of acrolein.}
    \label{fig:convergence_pol}
\end{figure}

\paragraph{Model of acrolein}
The parameterized model was obtained from Ref.~\citenum{TRUECARS_intro}, with some slight modifications. In particular, we added fourth-order terms in $q$ to the diagonal elements of the diabatic Hamiltonian, in order to ensure that the potential is bounded from below. This did not significantly change the potential in the region of interest, but was necessary since classical trajectories that escape to the unbounded region will diverge. This issue was not previously encountered in the full QM description due to the finite size of the grid. In the present work we use same the fourth-order diabatic Hamiltonian for all methods, and note that the full QM description leads only to slight differences with the previous work. The diabatic Hamiltonian is defined as:
\begin{align}
    \He = \begin{pmatrix}
        H_{\text{AA}}(q_1, q_2) & f_{\text{AB}}(q_1, q_2)g(q_1)\eu{-q_2^2 / 0.08} \\
        f_{\text{AB}}(q_1, q_2)g(q_1)\eu{-q_2^2 / 0.08} & H_{\text{AB}}(q_1, q_2)
    \end{pmatrix}
    \label{eq:Muk2_potential}
\end{align}
where
\begin{align*}
    g(q_1) &= \begin{cases}
        \eu{-q_1^2/0.18} \qquad\; q_1 < 0 \\
        \eu{-q_1^2/0.045} \qquad q_1 \geq 0 
    \end{cases}\\
    H_{\text{AA}}, f_{\text{AB}}, H_{\text{BB}} &= c_{00} + c_{10}q_1 + c_{01}q_2 + c_{20}q_1^2 + c_{11}q_1q_2 + c_{02}q_2^2 + c_{30}q_1^3 + c_{21}q_1^2q_2 + \\ & + c_{12}q_1q_2^2 + c_{03}q_2^3 + c_{40}q_1^4 + c_{04}q_2^4
\end{align*}
and all coefficients are reported in Table \ref{tab:cost_Muk2D}.
Figure~\ref{fig:2DMuk_pot} shows the diabatic energies ($H_\text{AA}$, $H_\text{AB}$), diabatic coupling $H_\text{BB}$, adiabatic energies ($V_0$, $V_1$) and nonadiabatic coupling $d = \langle\phi_0\vert \mathbf{\nabla} \vert\phi_1\rangle $. The dashed black lines indicate where the diabatic energies are equal, and the solid black lines indicate where the diabatic coupling is zero. Therefore the CIs are located at the two points where the black lines intersect.
\begin{table*}
    \centering
    \begin{tabular}{ld{2.7}d{2.7}d{2.7}d{2.7}d{2.7}d{2.7}d{2.7}}
        \toprule
        & \head{$c_{00}$} & \head{$c_{10}$} & \head{$c_{01}$} & \head{$c_{20}$} & \head{$c_{11}$} & \head{$c_{02}$} \\
        \midrule
        $H_{\text{AA}}$ & -0.01854  & -0.02817 & -0.114    &  0.3156  & -0.1576  & 0.2457  \\
        $f_{\text{AB}}$ & 0.0006653 & -0.05699 & -0.001481 & -0.02017 & -0.06204 & 0.02157 \\
        $H_{\text{BB}}$ & -0.001247 &  0.01804 & 0.02297   &  0.4546  & -0.2419  & 0.2242  \\
        \midrule
        & \head{$c_{30}$} & \head{$c_{21}$} & \head{$c_{12}$} & \head{$c_{03}$} & \head{$c_{40}$} & \head{$c_{04}$} \\
        \midrule
        $H_{\text{AA}}$ & 0.1237  & 0.2883  & -0.2856 & 0.1071  & 0.08 & 0.08 \\
        $f_{\text{AB}}$ & 0.06652 & 0.05527 & 0.04719 & 0.02031 & 0.0  & 0.0  \\
        $H_{\text{BB}}$ & 0.2404  & 0.1135  & -0.3448 & 0.07928 & 0.08 & 0.08 \\
        \bottomrule
    \end{tabular}
    \caption{Constants for the model of acrolein in atomic units.}
    \label{tab:cost_Muk2D}
\end{table*}%
The mass is $m = 30\,000$\,a.u., The initial nuclear wavepacket is a Gaussian on the excited state centered at $\bar{q} = (0.5, 0.5)$ with a spread corresponding to $\omega = 0.000125$ in both directions.
\begin{subequations}
\begin{align}
    \chi_1(q,t=0) &= \sqrt{\frac{m\omega}{\pi\hbar}}\,\eu{-\frac{m\omega}{2\hbar}(q_1-\bar{q}_1)^2}\,\eu{-\frac{m\omega}{2\hbar}(q_2-\bar{q}_2)^2} \\
    \rho_1(q_1, q_2, p_1, p_2, t=0) &= \frac{1}{\pi^2\hbar^2}\,\eu{-\frac{m\omega}{\hbar}(q_1-\bar{q}_1)^2}\,\eu{-\frac{m\omega}{\hbar}(q_2-\bar{q}_2)^2}\,\eu{-\frac{p_1^2}{m\hbar\omega}}\,\eu{-\frac{p_2^2}{m\hbar\omega}}
\end{align}
\end{subequations}
The timestep for the simulation is $\delta t = 0.02$ a.u. For MASH and FSSH the convergence has been ensured using $10^5$ trajectories, although almost perfect convergence is already obtained with $10^4$ trajectories and reasonable results are available from $10^3$ trajectories [Fig.~\ref{fig:convergence_pol}]. Note that both MASH and FSSH converge at the same rate.
%In both cases, the coherences converge slower than the populations.

The standard deviations for the Gaussian envelopes of the electric pulses are respectively $\Delta\omega_\mathrm{B} = 5$ and $\Delta t_\mathrm{N} = 48$ (both in atomic units).
\begin{subequations}
\begin{align}
    \mathcal{E}_\mathrm{B}^*(\omega) &= \eu{-\frac{\omega^2}{2 \Delta\omega_\mathrm{B}^2}} \\
    \mathcal{E}_\mathrm{N}(t) &= \eu{-\frac{t^2}{2\Delta t_\mathrm{N}^2}}
\end{align}
\end{subequations}
The Wigner spectrogram, $\mathcal{W}(T,\omega)$, is defined using a reference frequency of $\omega_\mathrm{R} = -0.05$.

\begin{figure*}
    \centering
    \includegraphics{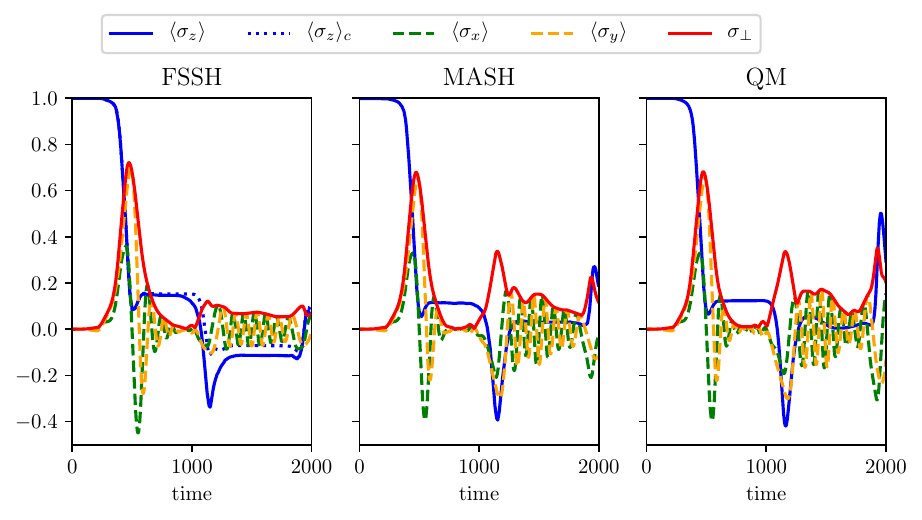}
    \caption{Adiabatic populations and coherences for the model of acrolein. For FSSH, we also plot $\braket{\sigma_z(t)}_c$ calculated from the coefficients $\braket{|c_1(t)|^2-|c_0(t)|^2}_\mathrm{FSSH}$. We also show the absolute value of the coherences $\sigma_\perp=\sqrt{\braket{\sigma_x}^2 + \braket{\sigma_y}^2}$ in red.}
    \label{fig:2DMuk_pop_coh}
\end{figure*}

\begin{figure*}[p]
    \centering
    \includegraphics[height=0.4\textheight]{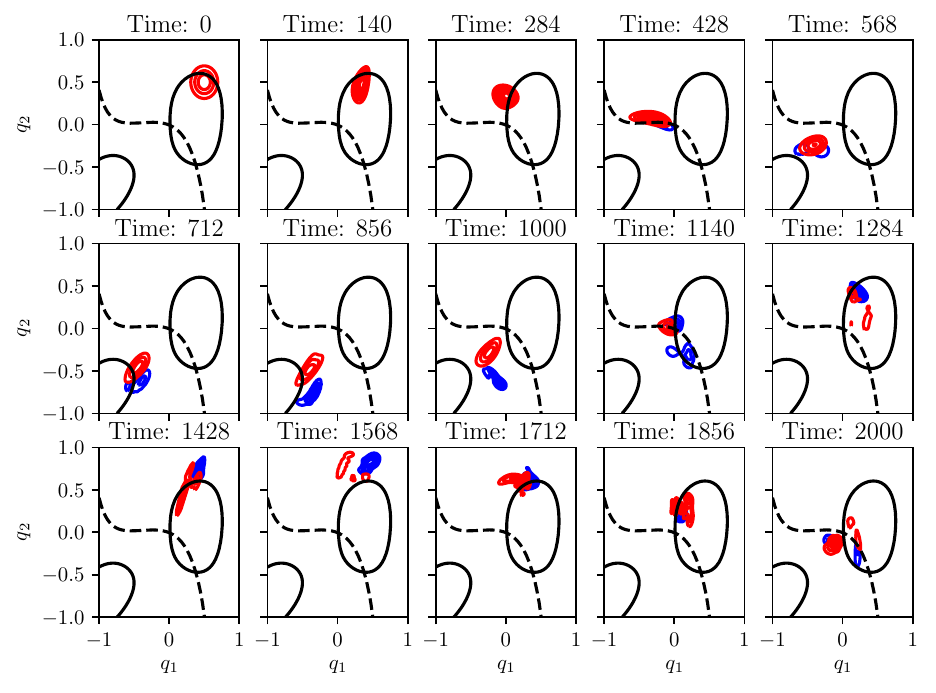}
    \caption{Quantum dynamics of the model of acrolein.  The nuclear density on the upper adiabatic state is plotted in red and on the lower adiabatic state in blue.}
    \label{fig:2DMuk_dens}
\end{figure*}

\begin{figure*}[p]
    \centering
    \includegraphics[height=0.4\textheight]{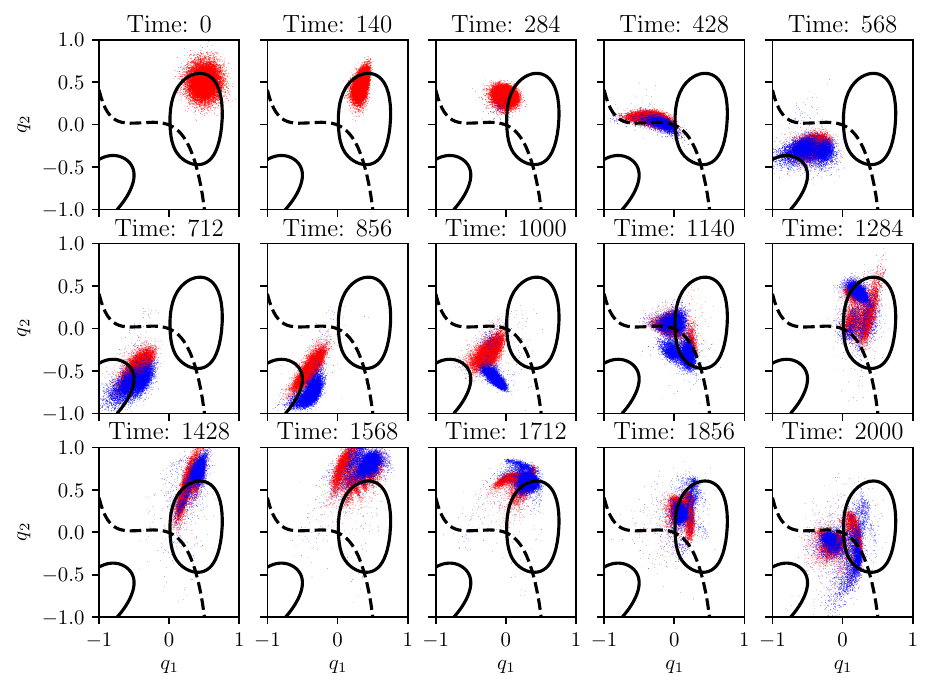}
    \caption{MASH dynamics of the model of acrolein.  The trajectories on the upper (lower) adiabatic state are red (blue) points.}
    \label{fig:2DMuk_dens_MASH}
\end{figure*}

To provide a simple interpretation of the TRUECARS signal from the various simulations, we employ the following idealizations 
to obtain the histogram of adiabatic energy gaps:
\begin{enumerate}
    \item Most of the time, the nuclear density is outside of coupling region, where
    \begin{align}
        \alpha_z(q)\approx 0, \qquad\alpha_x(q)\approx 1.
    \end{align}
    %in the regions of interest %(where $\chi_0(q)\chi_1(q)\neq 0$ for QM, $V_1(q(t))\neq V_0(q(t))$ for each trajectory).
    Therefore the time-dependent polarizability is given by
    \begin{align}
        \langle\alpha(t)\rangle \approx \langle\sigma_x(t)\rangle %\approx \langle\sigma_\perp(t)\cos(\omega_t t)\rangle
    \end{align}

\begin{comment}
    \item Next, we ignore the effect of phase interference between trajectories and write
    \begin{subequations}
    \begin{align}
        \mathrm{QM} & \qquad \langle\sigma_x(t)\rangle = \langle\sigma_\perp(t)\cos(\omega_t t)\rangle_q = 2\int\mathrm{d}q |\chi_1(q,t)\chi_0(q,t)|\cos(\Delta V(q)t) \\
        \mathrm{MASH} & \qquad \langle\sigma_x(t)\rangle = \langle\sigma_\perp(t)\cos(\omega_t t)\rangle_q = \langle\sqrt{\langle\sigma_x(t)\rangle^2_S+\langle\sigma_y(t)\rangle^2_S}\cos\Big[\Delta V(q(t)) t\Big]\rangle_q \\
        \mathrm{FSSH} & \qquad \langle\sigma_x(t)\rangle = \langle\sigma_\perp(t)\cos(\omega_t t)\rangle_q = 2\langle |\langle c_0c_1\rangle_c| \cos\Big[\Delta V(q(t)) t\Big]\rangle_q
    \end{align}
    \end{subequations}
    The $\langle\cdot\rangle_S$ and $\langle\cdot\rangle_c$ are averages over the electronic variables (spin vector or wave-function coefficients), at fixed position $q$
    and $\langle\cdot\range_q$ is an average over positions.
\end{comment}
    
    \item We assume that $\mathcal{E}_\mathrm{N}(t)$ is slowly decaying with time with respect to the frequencies $\omega_t=\Delta V(q(t))$
    but faster than the timescale of the nuclear dynamics.
    We thus approximate $\sigma_x(t)=\sigma_\perp(T)\sin(\omega_t t)$, where
    \begin{subequations}
    \begin{align}
        \mathrm{QM} & \qquad \langle\sigma_x(t)\rangle = 2\int\mathrm{d}q |\chi_1(q,t)\chi_0(q,t)|\sin(\Delta V(q)t) \\
        \mathrm{MASH} & \qquad \langle\sigma_x(t)\rangle = 2 \left\langle h(S_z(0)) \, \sqrt{S_x(t)^2+S_y(t)^2}\sin\Big[\Delta V(q(t)) t\Big]\right\rangle_\text{MASH} \\
        \mathrm{FSSH} & \qquad \langle\sigma_x(t)\rangle = 2\left\langle |\langle c_0(t)c_1(t)\rangle| \sin\Big[\Delta V(q(t)) t\Big]\right\rangle_\text{FSSH}
    \end{align}
    \end{subequations}

    \item We assume that $\mathcal{E}_\mathrm{B}^*(\omega)$ is so broad as to be effectively uniform for all relevant frequencies.
\end{enumerate}

The histogram of the signal is thus reduced to:
\begin{subequations}
\begin{align}
    P(\omega',T) &= \int_0^\infty  \mathrm{d}\omega \, \mathcal{S}(\omega, T) \, \delta(\omega-\omega') \\
    &= \int_0^\infty \mathrm{d}\omega \, 2\Im\int_{-\infty}^{+\infty} \mathrm{d}t\, \eu{\iu\omega(t-T)}\mathcal{E}_\mathrm{B}^*(\omega)\mathcal{E}_\mathrm{N}(t-T) \langle\sigma_x(t)\rangle \, \delta(\omega-\omega') \\
    &\approx \int_0^\infty \mathrm{d}\omega \, 2\Im\int_{-\infty}^{+\infty} \mathrm{d}t\, \eu{\iu\omega(t-T)} \mathcal{E}_\mathrm{N}(t-T) \langle\sigma_\perp(T)\sin(\omega_t t) \rangle \, \delta(\omega-\omega') \\
    &\approx \int_0^\infty \mathrm{d}\omega \, 2\pi \langle\sigma_\perp(T) \delta(\omega-\omega_t) \rangle \, \delta(\omega-\omega') \\
    &= 2\pi \langle\sigma_\perp(T) \delta(\omega-\omega') \rangle
\end{align}
\end{subequations}
%\begin{align}
%    \int_0^\infty  \mathrm{d}\omega \, \mathcal{S}(\omega, T) \propto \langle 2 \mathcal{E}_\mathrm{B}^*(\omega_T) \sigma_\perp(T) \rangle_q
%\end{align}
Therefore the average shift is defined as
\begin{align}
    \bar{\omega}(t) = \frac{\langle \sigma_\perp(t) \, \omega_t \rangle}{\langle \sigma_\perp(t) \rangle}
\end{align}
%The last approximation for the trajectory-based methods MASH and FSSH is that the average on spin-vector (coefficients) can be moved outside of the square root (absolute value).
or more explicitly for each method:
\begin{subequations}
\label{eq:sum_average_shift}
\begin{align}
    \mathrm{QM} & \qquad \bar{\omega}(t) = \frac{\int\mathrm{d}q\, |\chi_1(q,t)\chi_0(q,t)|\,\Delta V(q) }{\int\mathrm{d}q\, |\chi_1(q,t)\chi_0(q,t)|} \\
    \mathrm{MASH} & \qquad \bar{\omega}(t) = \frac{\langle h(S_z(0))\,\sqrt{S_x(t)^2 + S_y(t)^2} \, \Delta V(q(t))\rangle_\text{MASH}}{\langle h(S_z(0))\,\sqrt{S_x(t)^2 + S_y(t)^2} \rangle_\text{MASH}} \\
    \mathrm{FSSH} & \qquad \bar{\omega}(t) = \frac{\langle |c_1(t)c_0(t)| \,\Delta V(q(t))\rangle_\text{FSSH}}{\langle |c_1(t)c_0(t)| \rangle_\text{FSSH}}
\end{align}
\end{subequations}

\begin{figure*}
    \centering
    \includegraphics{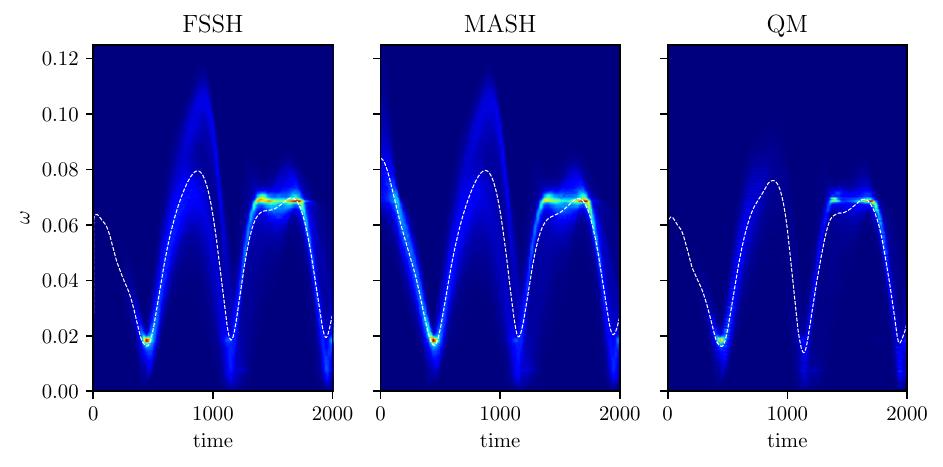}
    \caption{Energy shift between the adiabatic states $\Delta V$, weighted by the corresponding factor in [Eq.~\eqref{eq:sum_average_shift}]: $|\chi_1(q,t)\chi_0(q,t)|$, $\sqrt{S_x(t)^2 + S_y(t)^2}$ and $|c_1(t)c_0(t)|$. In each case, we also plot the mean frequency $\bar{\omega}(T)$ as a white dashed line. Note that these definitions are not true physical observables; they are just useful to interpret the signal.}
    \label{fig:mean_frequency}
\end{figure*}

\end{document}